\documentclass[11pt]{article}
\newcommand{\bm}[1]{\mbox{\boldmath $#1$}}

\def\ds{\displaystyle} \def\bb{\bibitem} \def\lb{\label}
\def\be{\begin{equation}} \def\ee{\end{equation}}
\def\ba{\begin{eqnarray}} \def\ea{\end{eqnarray}} \def\part{\partial}
\def\nn{\nonumber}
\def\ol{\overline}  \def\k{\kappa} \def\L{\Lambda}
\def\X{{\bm X}} \def\S{{\bm S_E}}
\def\a{{\bm \alpha}} \def\b{{\bm \beta}} \def\c{{\bm \gamma}}
 \def\J{{\bm J}}

\def\bL{{\bm L}}

\begin{document}
\begin{titlepage}
\date{26 July 2008}

\title{
\begin{flushright}\begin{small}    LAPTH-1262/08
\end{small} \end{flushright} \vspace{1.5cm}
Three-dimensional Chern-Simons black holes}

\author{Karim Ait Moussa$^a$ \thanks{Email: karim.aitmoussa@wissal.dz},
G\'erard Cl\'ement$^b$ \thanks{Email: gclement@lapp.in2p3.fr}, Hakim
Guennoune$^{b,c}$ \thanks{Email: guenoun@lapp.in2p3.fr}, C\'edric
Leygnac$^b$ \thanks{Email: leygnac@lapp.in2p3.fr}  \\ \\
{$^a$ \small Laboratoire de Physique Math\'ematique et Physique
Subatomique, D\'epartement de Physique,} \\{\small Facult\'e des
Sciences, Universit\'e Mentouri, Constantine 25000, Algeria} \\
{$^b$ \small Laboratoire de  Physique Th\'eorique LAPTH (CNRS),
B.P.110, F-74941 Annecy-le-Vieux cedex, France}\\
{$^c$ \small D\'epartement de Physique, Facult\'e des Sciences,
Universit\'e Ferhat Abbas, S\'etif 19000, Algeria}}

\maketitle

\begin{abstract}
We construct black hole solutions to three-dimensional
Einstein-Maxwell theory with both gravitational and electromagnetic
Chern-Simons terms. These intrinsically rotating solutions are
geodesically complete, and causally regular within a certain
parameter range. Their mass, angular momentum and entropy are found
to satisfy the first law of black hole thermodynamics. These
Chern-Simons black holes admit a four-parameter local isometry
algebra, which generically is $sl(2,R)\times R$, and may be
generated from the corresponding vacua by local coordinate
transformations.

\end{abstract}

\end{titlepage}
\setcounter{page}{2}

\section{Introduction}

Three-dimensional gravity admits a variety of black hole solutions.
The first discovered, and most well-known, is the
Ba\~{n}ados-Teitelboim-Zanelli (BTZ) black hole solution to
three-dimensional gravity with a negative cosmological constant
\cite{btz}. Black holes in a generalized three-dimensional dilaton
gravity theory were analysed in \cite{skl}. Three discrete families,
each classified by an integer, of black hole solutions to the theory
of a massless scalar field coupled repulsively to
gravity\footnote{In contrast to the four-dimensional case, the
gravitational constant can be either positive or negative in three
dimensions \cite{djh}.} were found in \cite{sig}. More recently, it
was shown in \cite{tmgbh} that topologically massive gravity (TMG)
\cite{djt} --Einstein gravity augmented by a gravitational
Chern-Simons term-- with a vanishing cosmological constant admits
non-asymptotically flat (and non-asymptotically AdS), intrinsically
non-static black hole solutions. The computation of the mass and
angular momentum of these ACL black holes presented a challenge
which was successively addressed and solved in \cite{adtmg}.

The purpose of the present work is to construct and analyse black
hole solutions to topologically massive gravitoelectrodynamics
(TMGE), three-dimensional Einstein-Maxwell theory augmented by both
gravitational and electromagnetic Chern-Simons terms. In
\cite{tmge}, two classes of exact solutions of this theory were
obtained by making suitable ans\"atze, the first leading to
geodesically complete self-dual stationary solutions, and the other
to diagonal solutions, including black point static solutions and
anisotropic cosmologies. We will show in the following that a third,
simple ansatz yields black hole solutions generalizing those of
\cite{tmgbh}.

In the next section, we introduce the model, and summarize the
dimensional reduction procedure followed in \cite{tmge} to derive
the field equations for solutions with two Killing vectors. Our
ansatz leads in the third section to three black hole solutions
(according to the values of the model parameters) depending
generically on two integration constants. In the fourth section we
analyse the global structure of our black hole spacetimes. After
analytical extension through the two horizons, these are
geodesically complete, but may allow closed timelike curves in
certain parameter domains. We show that in the latter case it is
possible to further narrow the parameter range so that the acausal
region is hidden behind the event horizon. The mass, angular
momentum and entropy of these black holes, computed in the fifth
section, are checked to satisfy in all cases the first law of black
hole thermodynamics for independent variations of the black hole
parameters, as well as an integral Smarr-like relation. Finally, we
show in the sixth section that our black hole metrics admit four
local Killing vectors generating either the $sl(2,R)\times R$
algebra or (in a special case) a solvable Lie algebra. The existence
of these four local isometries suggests that for a given set of
model parameters the black hole solutions depending on different
integration constants may be transformed into each other by local
coordinate transformations, which we give explicitly. We close with
a brief discussion.

\section{The model}
The action for TMGE may be written
\be \lb{act}
I = I_E + I_M + I_{CSG} + I_{CSE}\,,
\ee
where
\ba
I_E & = & \frac1{2\k}\int d^3x \sqrt{|g|}\, (R-2\L)\,,  \nonumber \\
I_M & = & -\frac{1}{4} \int d^3x \sqrt{|g|}\, g^{\mu\nu}g^{\rho\sigma}F_{\mu
\rho}F_{\nu\sigma} \,.
\ea
are the Einstein action (with cosmological constant $\L$ and
Einstein gravitational constant $\k = 8\pi G$) and the Maxwell
action,  and
\ba
I_{CSG} & = & \frac{1}{4\k\mu_G} \int d^3x \,\epsilon^{\lambda\mu\nu}\,
\Gamma^\rho_{\lambda\sigma} \left[\partial_\mu
\Gamma_{\rho\nu}^\sigma+\frac{2}{3}\,\Gamma^\sigma_{\mu\tau}
\Gamma_{\nu\rho}^\tau \right]\,,  \nonumber \\
I_{CSE} & = &
\frac{\mu_E}{2} \int d^3x \,\epsilon^{\mu\nu\rho} A_\mu
\partial_\nu A_\rho\,,
\ea
are the gravitational and electromagnetic Chern-Simons terms
($\epsilon^{\mu\nu\rho}$ is the antisymmetric symbol), with Chern-Simons
coupling constants $1/\mu_G$ and $\mu_E$.

We shall search for stationary circularly symmetric solutions of this
theory using the dimensional reduction procedure of \cite{tmge}, which
we summarize here. We choose the parametrisation \cite{EL,EML}
\be \lb{par}
ds^2=\lambda_{ab}(\rho)\,dx^a dx^b + \zeta^{-2}(\rho)R^{-2}(\rho) \,d\rho^2\,,
\qquad A = \psi_a(\rho) \,dx^a
\ee
($x^0 = t$, $x^1 = \varphi$), where $\lambda$ is the $2 \times 2$ matrix
\be
\lambda = \left(
\begin{array}{cc}
T+X & Y \\
Y & T-X
\end{array}
\right),
\ee
$R^2 \equiv \X^2$ is the Minkowski pseudo-norm
of the ``vector'' $\X(\rho) = (T,\,X,\,Y)$,
\be
\X^2 = \eta_{ij}\,X^iX^j = -T^2+X^2+Y^2 \,,
\ee
and the scale factor $\zeta(\rho)$ allows for arbitrary reparametrizations
of the radial coordinate $\rho$. We recall for future purpose that
stationary solutions correspond to ``spacelike'' paths $\X(\rho)$ with
$R^2 > 0$, and that intersections of these paths with the future light cone
($R^2 = 0$, $T > 0$) correspond to event horizons.

The parametrization (\ref{par}) reduces the action (\ref{act}) to the form
\be
I = \int d^2x \int d\rho \, L \,,
\ee
with the effective Lagrangian $L$
\ba \lb{L1}
L & = & \frac{1}{2}\bigg[ \frac{1}{2\k\mu_G}\,\zeta^2
\X\cdot(\dot{\X} \wedge \ddot{\X}) + \frac1{2\k}\,\zeta \
\dot{\X}^2  \nonumber \\
& &  + \zeta\,\dot{\ol\psi}\,\bm{\Sigma} \cdot \X\,\dot{\psi} +
\mu_E\,\ol{\psi}\,\dot{\psi} - \frac2{\k}\,\zeta^{-1}\L \bigg] \,.
\ea In (\ref{L1}), $\dot{} = \partial/\partial\rho$, the wedge
product is defined by $({\bf X} \wedge {\bf Y})^i =$
$\eta^{ij}\epsilon_{jkl}X^k Y^l$ (with $\epsilon_{012} = +1$), the
``Dirac'' matrices $\Sigma^i$ are \be \lb{Dirac} \Sigma ^0 = \left(
\begin{array}{cc}
0 & 1 \\
-1 & 0
\end{array}
\right) \, , \,\,\,
\Sigma ^1 = \left(
\begin{array}{cc}
0 & -1 \\
-1 & 0
\end{array}
\right) \, , \,\,\,
\Sigma ^2 = \left(
\begin{array}{cc}
1 & 0 \\
0 & -1
\end{array}
\right) \, ,
\ee
and $\ol{\psi} \equiv \psi^T\,\Sigma^0$ is the (real) Dirac adjoint of the
``spinor'' $\psi$.

Variation of the Lagrangian (\ref{L1}) with respect to
$\psi$ leads to the equation
\be\lb{CSeq}
\part_{\rho}\bigg[\zeta(\bm{\Sigma}\cdot\X)\dot{\psi} + \mu_E\psi\bigg] = 0\,.
\ee
This means that the bracket is a constant of the motion, which may be set
to zero by a gauge transformation, leading to the first integral
\be \lb{psi}
\zeta\dot{\psi} = \frac{\mu_E}{R^2}\, \bm{\Sigma} \cdot
\X\, \psi\,.
\ee
This allows to eliminate from the Lagrangian (\ref{L1}) the spinor
fields $\psi$ and $\dot{\psi}$ in favor of the null ($\S^2 = 0$)
``spin'' vector field
\be\lb{spin}
\S = -\frac{\k}{2}\,\ol{\psi} \bm{\Sigma} \psi \,,
\ee
satisfying the equation (equivalent to (\ref{psi}))
\be\lb{S}
\zeta\dot{\S}  = \ds\frac{2\mu_E}{R^2} \X\wedge\S\,.
\ee
Varying the Lagrangian (\ref{L1}) with respect to $\X$, we then obtain
the dynamical equation for the vector fields $X$,
\ba \lb{X}
\ddot{\X}  & = & \ds\frac{\zeta}{2\mu_G} \bigg[\,3(\dot{\X} \wedge
\ddot{\X}) +2(\X \wedge \dot{\ddot{\X}})\,\bigg] \nonumber\\
&   & -\ds\frac{2\mu_E^2}{\zeta^2R^2} \bigg[\S - \ds\frac{2}{R^2}
\X(\S \cdot \X)\bigg]\,, \ea where for simplicity we have fixed the
scale $\zeta =$ constant. Finally, variation of the Lagrangian
(\ref{L1}) with respect to the Lagrange multiplier $\zeta$ leads to
the Hamiltonian constraint \be \lb{Ham} H \equiv \frac{1}{4\k}
\left[ \dot{\X}^2 + 2\X \cdot \ddot{\X} - \frac{\zeta}{\mu_G} \X
\cdot (\dot{\X} \wedge \ddot{\X}) + 4\frac{\L}{\zeta^2} \right] =
0\,, \ee where we have used the equation (following from (\ref{X}))
\be\lb{SX} \S\cdot\X =
\frac{\zeta^2R^2}{2\mu_E^2}\bigg[\X\cdot\ddot{\X} -
\frac{3\zeta}{2\mu_G}\X\cdot(\dot{\X} \wedge \ddot{\X})\bigg]\,. \ee

In the preceding equations we have kept the constant scale parameter
$\zeta$ free. From the circularly symmetric ansatz (\ref{an}) we find
det$|g| = \zeta^{-2}$, showing that $\zeta$ has the same dimension
(an inverse length) as the Chern-Simons coupling constants $\mu_G$ and
$\mu_E$. In the following it will prove convenient to fix the scale to
\be
\zeta = \mu_E\,
\ee
(assumed, without loss of generality, to be positive), and to trade the
gravitational Chern-Simons coupling constant $\mu_G$ for the dimensionless
parameter
\be
\lambda \equiv \frac{\mu_E}{2\mu_G}\,.
\ee

\section{Black hole solutions}
\setcounter{equation}{0}

Two special classes of solutions to the reduced field equations
(\ref{psi})-(\ref{Ham}) of
TMGE have been given in \cite{tmge}: self-dual solutions, which are
asymptotically Minkowski or anti-de Sitter, and (in the case of exact
balance $\mu_G + \mu_E = 0$) diagonal (static) solutions. In this
paper, we shall derive the class of black hole solutions from the
ansatz \cite{part}
\be\lb{an}
\X =\a\,\rho^2 + \b\,\rho + \c\,,
\ee
with $\a$, $\b$ and $\c$ linearly independent constant vectors.

We first insert this ansatz in the Hamiltonian constraint (\ref{Ham}),
which reduces to a second order equation in $\rho$. This is
identically satisfied, provided the vectors $\a$, $\b$ and $\c$ are
constrained by
\ba
&& \a^2 = 0\,, \qquad (\a\cdot\b) = 0\,, \lb{null}\\
&& \b^2 + 4\bigg[\a\cdot\c + \lambda(\a\wedge\b)\cdot\c +
\frac{\L}{\mu_E^2}\bigg] = 0\,. \lb{cons}
\ea
It is easy to show that the two constraints (\ref{null}) further imply
\be\lb{d}
\a\wedge\b = d\a\,, \qquad \b^2 = d^2\,,
\ee
for some constant $d$. Next, we compute from (\ref{SX})
\be
\S\cdot\X = b(\a\cdot\c)R^2\,, \qquad b =
1+3d\lambda\,.
\ee
Inserting this into (\ref{X}), we obtain
\be\lb{Sb}
\S = b\bigg[2(\a\cdot\c)\X - \a R^2\bigg]\,.
\ee
Differentiating this, we obtain after some algebra
\ba\lb{dotS}
\dot{\S} &=& 2b\bigg[-(\b^2)\a\rho + (\a\cdot\c)\b -
  (\b\cdot\c)\a\bigg]
\nonumber \\
&=& -2bd\bigg[d\a\rho + \a\wedge\c\bigg]\,,
\ea
where the wedge product $\a\wedge\c$ has been evaluated with the aid
of (\ref{d}). On the other hand,
\be\lb{XwS}
\X\wedge\S = b\bigg[d\a\rho + \a\wedge\c\bigg]R^2\,.
\ee
Comparing Eqs. (\ref{dotS}) and (\ref{XwS}), we see that Eq.
(\ref{S}) for the gauge field is satisfied if either $b=0$, leading to
$d=-2\mu_G/3\mu_E$ (this is the case of TMG), or
\be
d = -1\,.
\ee
In this case the remaining constraint equation (\ref{cons}) then
leads to the value of the scalar product $\a\cdot\c$
\be\lb{ac}
\a\cdot\c = -\frac{1+4\L/\mu_E^2}{4(1-\lambda)}\,.
\ee

At this stage the solution depends on five parameters (nine parameters in
the ansatz (\ref{an}) restricted by the four constraints $\a^2 = 0$,
$\a\wedge\b = -\a$ (two constraints), and (\ref{ac})). In principle the
number of free parameters may be reduced to two by taking into account
the three-parameter group of transformations which leave the ansatz
(\ref{an}) (with $\zeta = \mu_E$ and the period of the angle $\varphi$
fixed) form-invariant: translations of $\rho$, transition to uniformly
rotating frames, and simultaneous length ($\rho$) and time ($t$)
rescalings. However, for the sake of comparison with \cite{BBCG}, we
will allow for an arbitrary time scale parameter $\sqrt{c}$. Choose a
rotating frame and a
length-time scale such that $\a = (c/2,-c/2,0)$. In this frame,
\be
\a = (c/2,-c/2,0)\,,\quad \b = (\omega,-\omega,-1)\,, \quad \c =
(z+u,z-u,v)\,,
\ee
with $\a\cdot\c = -cz$ given by (\ref{ac}),
\be\lb{z}
z = (1-\beta^2)/2c,
\ee
where
\be\lb{beta}
\beta^2 = \frac{1-2\lambda-4\L/\mu_E^2}{2(1-\lambda)}\,.
\ee
Computation of $R^2$ gives
\be\lb{R2}
R^2 = \beta^2\rho^2 - 2(v+ 2\omega z)\rho + v^2 - 4uz\,.
\ee
If $\beta^2\neq0$ the linear term can be set to zero ($v = -2\omega
z$) by a translation of $\rho$, leading to
\be\lb{R2b}
R^2 = \beta^{2}(\rho^2-\rho_0^2)\,,
\ee
where we have eliminated $u$ in terms of $z$ and the new real parameter
$\rho_0^2$. The final
metric may be written in the two equivalent forms:
\ba
ds^2 &=&
\frac{1-\beta^2}c\bigg[dt-\bigg(\frac{c\rho}{1-\beta^2}+\omega\bigg)\,
d\varphi\bigg]^2
- \frac{c\beta^2}{1-\beta^2}(\rho^2-\rho_0^2)\,d\varphi^2 \nonumber \\ &&\qquad
+  \frac1{\beta^2\mu_E^2}\frac{d\rho^2}{\rho^2-\rho_0^2}\,, \lb{bh1}
\ea
and
\ba
ds^2 &=& -\beta^2\frac{\rho^2-\rho_0^2}{r^2}\,dt^2 + r^2\bigg[d\varphi
  - \frac{\rho+(1-\beta^2)\omega/c}{r^2}\,dt\bigg]^2 \nonumber \\ &&\qquad
+ \frac1{\beta^2\mu_E^2}\frac{d\rho^2}{\rho^2-\rho_0^2}\,, \lb{bh2}
\ea
with
\be\lb{r2}
r^2 = c\rho^2 +2\omega\rho + \frac{\omega^2}c\,(1-\beta^2) +
\frac{c\beta^2\rho_0^2}{1-\beta^2}\,.
\ee

This metric is very similar in form to that of \cite{tmgbh}. While a
superficial glance at (\ref{bh1}) would suggest that the metric is
Lorentzian provided $\beta^2 > 1$ for $c>0$, or $\beta^2 < 1$ for
$c<0$, the ADM form (\ref{bh2}) shows that the condition $\beta^2 >
0$ is sufficient. Obviously this is a black hole, with two horizons
at $\rho = \pm \rho_0$, if $\rho_0^2 > 0$. The analysis of the
causal structure of this spacetime, carried out in the next section,
further shows that this is a causally regular black hole provided
$c>0$ and $\beta^2 < 1$. So the range of regularity of this black
hole solution is \be 0< \beta^2 < 1 \quad \Leftrightarrow \quad
\left|
\begin{array}{ccc} 2\lambda > 1 - \frac{4\L}{\mu_E^2}
 & {\mbox{if}} & \L < -\frac{\mu_E^2}4 \,, \\
2\lambda < 1 - \frac{4\L}{\mu_E^2}
& {\mbox{if}} & \L > -\frac{\mu_E^2}4 \,.
\end{array} \right.
\ee

The form of the metric breaks  down for $\beta^2 =
1$ ($\L = - \mu_E^2/4$) and $\beta^2 = 0$ ($2\lambda = 1 -
4\L/\mu_E^2$), as well as in the intersection ($\lambda = 1$ with
$\L = -\mu_E^2/4$) of these two cases. However it is possible to
extend the solution to all these cases. We first consider the case
$\underline{\beta^2 = 1}$. In this case the parameter $z = 0$, and so
also $v = 0$, and $\rho_0 = 0$, with the parameter $u$ remaining free. So the
metric in this case is (\ref{bh2}) with
\be\lb{r21}
\beta^2 = 1, \quad \rho_0 = 0, \quad r^2 = c\rho^2 + 2\omega\rho + 2u\,.
\ee

In the case $\underline{\beta^2 = 0}$ ($z=1/2c$), the
quadratic term is absent from (\ref{R2}). If $v+\omega/c \neq 0$
the constant term can be set
to zero ($u = cv^2/2$) by a translation of $\rho$, so that (\ref{R2b})
is replaced by
\be
R^2 = 2\nu\rho\,,
\ee
with $\nu = -v-\omega/c > 0$. The metric may be written as
\be
ds^2 = \frac1c\bigg[dt-c(\rho+\nu+\omega/c)\,d\varphi\bigg]^2 -
2c\nu\rho\,d\varphi^2  + \frac{d\rho^2}{2\mu_E^2\nu\rho}\,, \lb{bh01}
\ee
or
\be
ds^2 = -\frac{2\nu\rho}{r^2}\,dt^2 + r^2\bigg[d\varphi
  - \frac{\rho+\nu+\omega/c}{r^2}\,dt\bigg]^2 +
\frac{d\rho^2}{2\mu_E^2\nu\rho}\,, \lb{bh02}
\ee
with
\be\lb{r20}
r^2 = c\rho^2 +2\omega\rho + c(\nu+\omega/c)^2\,.
\ee
This is a causally regular black hole, with a single horizon at $\rho
= 0$, if $c>0$ and $\omega > -c\nu/2$. The formal limit $\nu \to 0$ corresponds
to the exceptional case $v+\omega/c=0$, with the horizonless metric (in
which we have translated $\rho \to \rho - \omega/c$, and put $2u \equiv
-c\rho_1^2$)
\ba
ds^2 &=& \frac1c\bigg[dt-c\rho d\varphi\bigg]^2 -c\rho_1^2\,d\varphi^2 +
\frac{d\rho^2}{\mu_E^2\rho_1^2} \lb{vac01}\\
&=& -\frac{\rho_1^2}{c(\rho^2-\rho_1^2)}\,dt^2 +
c(\rho^2-\rho_1^2)\bigg[d\varphi -
  \frac{\rho}{c(\rho^2-\rho_1^2)}\,dt\bigg]^2 +
\frac{d\rho^2}{\mu_E^2\rho_1^2}\,.\lb{vac02}
\ea

Finally, in the case $\underline{\lambda = 1}$ ($\mu_E/\mu_G = 2)$
with $\underline{\L=
  -\mu_E^2/4}$, the constraint (\ref{cons}) is {\em identically} satisfied
for all values of the scalar product $\a\cdot\c = -z$, i.e. for all
values of $\beta^2$. So the regular black hole solution is again given
by the generic forms (\ref{bh1}) or (\ref{bh2}) for $0 <
\beta^2 \le 1$ (with $\rho_0=0$ for $\beta^2=1$), where now $\beta^2$
is a {\em free} parameter, or by
(\ref{bh01}) or (\ref{bh02}) in the limiting case $\beta^2 = 0$.

The electromagnetic field generating this gravitational field may be
obtained by solving Eq. (\ref{spin}) for $\psi$, with $\S$ given by
(\ref{Sb}). This gives
\be\lb{Agen}
A = \pm\sqrt{-\frac{c\,(1-3\lambda)}{\k}}
\bigg[\frac{(1-\beta^2)}cdt-(\rho+(1-\beta^2)\omega/c)\,d\varphi\bigg]
\ee
for $0<\beta^2 \le 1$, or
\be\lb{A0}
A = \pm\sqrt{-\frac{c\,(1-3\lambda)}{\k}}
\bigg[c^{-1}dt-(\rho+\nu+\omega/c)\,d\varphi\bigg]
\ee
for $\beta^2 = 0$. Clearly this field is purely magnetic, as the
constant electric potential can be set to zero by a gauge transformation.

The reality of this electromagnetic potential
leads to an additional restriction on the
domain of existence of these black hole solutions. Recall that in
(2+1)-dimensional Einstein gravity, the sign of the gravitational
constant $\k$ is not fixed {\em a priori} \cite{djh}. It has been
argued \cite{djt} that in the case of topologically massive gravity
the gravitational constant should be taken negative to avoid the
appearance of ghosts. This argument follows from considering the TMG
action linearized around Minkowski spacetime. However, similarly to
the case of the BTZ black hole solutions of (2+1)-dimensional gravity
with a negative cosmological constant \cite{btz}, it might be more
appropriate to linearize the TMGE action around a suitable ``vacuum''
solution belonging to the black hole family, and it is not clear in
this case what sign of $\k$ should be taken. We shall consider both
signs to be possible. For $\underline{\k c > 0}$, the ratio of the two
Chern-Simons coupling constants must be bounded by
$\underline{\mu_E/\mu_G > 2/3}$. This leads to regular black holes
($0 \le \beta^2 \le 1$) only if $c>0$ and $\L < \mu_E^2/12$. For
$\underline{\k c < 0}$, the bound is inversed, $\underline{\mu_E/\mu_G
< 2/3}$. This leads to regular black holes if $c>0$ and $\L \ge -\mu_E^2/4$.

Finally the electromagnetic field vanishes for
$\underline{\mu_E/\mu_G = 2/3}$. So, when the two Chern-Simons
coupling constants are constrained by this relation, the black hole
metrics (\ref{bh1}) or (\ref{bh2}) (or (\ref{bh01}) or (\ref{bh02})
for $\L = \mu_E^2/12 = \mu_G^2/27$) again solve the equations of TMG.
For $c<0$, these $\lambda = 1/3$
solutions coincide (after appropriate coordinate transformations) with
the black hole solutions to TMG with cosmological constant given in
\cite{nutku,gurses} (see also \cite{part}, Eq. (18)). Regular black
holes (with $c>0$) were not correctly identified in these papers. For
$\L=0$ ($\beta^2 = 1/4$) and $c=1/4$, the black hole metrics (\ref{bh1})
and (\ref{bh2}) reduce respectively, after rescaling $\rho \to 2\rho$,
$\rho_0 \to 2\rho_0$ and in
units such that $\mu_G = 3$, to Eqs. (4) and (6) of \cite{tmgbh}.

The electromagnetic field also vanishes in the exceptional case $c=0$,
corresponding to $\a = 0$. In this case the metric \cite{EL}
$\X = \b\rho + \c$, with $\b^2 = -4\L/\mu_E^2$ from (\ref{cons}),
describes for $\L < 0$ the well-known BTZ black holes \cite{btz}.

To close this section we comment on the relation between our black
hole solutions and those of \cite{BBCG}. In the limit $\lambda \to
0$ ($\mu_G \to \infty$), the gravitational Chern-Simons coupling
constant goes to zero, and TMGE reduces for a negative cosmological
constant $\Lambda = -\ell^{-2}$ to the Einstein-Maxwell-Chern-Simons
theory considered in \cite{BBCG}, with $\alpha = -\mu_E/2$. For $\k
= 8\pi G$ positive (as assumed in \cite{BBCG}), the electromagnetic
field (\ref{Agen}) can be real only if $c<0$, so that the solutions
necessarily admit naked closed timelike curves. Noting that our
constant $\beta^2$ is given for $\mu_G \to \infty$ by \be 1-\beta^2
= \frac{\alpha^2\ell^2-1}{2\alpha^2\ell^2}, \ee we find that the
``G\"{o}del cosmon'' solution ((17) of \cite{BBCG}) with
$\alpha^2\ell^2
> 1$ and the ``G\"{o}del black hole'' solution ((31) of \cite{BBCG})
can both be put in the form (\ref{bh1}) or (\ref{bh2}), where our
radial coordinate $\rho$ is related to the radial coordinate $\ol r$
of \cite{BBCG} by $\rho = 2\alpha\ol{r} - (1-\beta^2)\omega/c$, with
$c=\mp(1-\beta^2)$ (upper sign in the case of (17) and lower sign in
the case of (31)), and our integration constants $\omega$ and
$\rho_0$ are related to the constants $\nu$ and $J$ by \be \omega =
\Gamma\nu\,, \quad \rho_0^2 = \Gamma^2\nu^2 \pm 2\Gamma J \qquad
\bigg(\Gamma = \frac{2G}{\alpha\beta^2} = \frac{4\alpha
\ell^2G}{1+\alpha^2\ell^2}\bigg)\,. \ee

\section{Global structure}
\setcounter{equation}{0}

Although the black hole spacetimes obtained in the previous section
are not  constant curvature, their curvature invariants are constant,
\ba\lb{curv}
{\cal R} &=& \frac{1-4\beta^2}2\mu_E^2\,, \nn\\ {\cal R}_{\mu\nu}R^{\mu\nu} &=&
\frac{3-8\beta^2+8\beta^4}4\mu_E^4\,, \ea
and depend only on the
parameter $\beta^2$. These spacetimes are clearly regular for all
$\rho \neq \pm\rho_0$, and may (except in the special case  $\omega =
\rho_0/\sqrt{1-\beta^2}$, see below) be extended through the horizons
$\rho = \pm\rho_0$ by the usual Kruskal method, leading to
geodesically complete spacetimes. However they may have  closed
timelike curves (CTC) for certain ranges of their parameters.  The
circles $\rho =$ constant are timelike whenever  $g_{\varphi
\varphi}=r^2<0$. For $c<0$, $r^2$ is negative at spacelike infinity
($\rho \to \infty$), so these solutions always admit CTC outside the
outer horizon. To exclude these we will assume in the  following
$c>0$. Fixing the scale so that $c = +1$, we find that in the  case of
the generic black hole (\ref{bh2}), the zeroes of $r^2$ are located at
\be \rho_{\pm}=-\omega \pm \beta\sqrt{\omega^2
-\frac{{\rho_0}^2}{1-\beta^2}} \, .  \ee It follows that CTC are
absent for $\beta^2 < 1$ and \be \omega^2<{\rho_0}^2/(1-\beta^2)\,.
\ee The Penrose diagram for the maximally extended spacetime is then
similar  to that of the Kerr black hole. For $\beta^2 < 1$ and
$\omega^2> {\rho_0}^2/(1-\beta^2)$, CTC do occur in the range $\rho
\in  [\rho_-,\rho_+]$, but are hidden behind the two horizons
($\rho_- < \rho_+ < -\rho_0$) if $\omega > 0$. The Penrose diagram
with the acausal regions cut out is the same as for
Reissner-Nordstr\"{o}m black holes. For other values of $\beta$ or
$\omega$, CTC occur outside the outer horizon. The limiting case
$\rho_0 = 0$ corresponds to extreme black holes, with an acausal
region behind the horizon for $\omega > 0$. The corresponding Penrose
diagram (again with the acausal regions cut out) is similar to that of
extreme Reissner-Nordstr\"{o}m black holes. In the exceptional case
$\omega = \rho_0/(1-\beta^2)$, the inner horizon $-\rho_0$ coincides
with the outer boundary $\rho_+$ of the acausal region, with the
metric reducing to \ba\lb{sch} ds^2 &=&
-\beta^2\frac{\rho-\rho_0}{\rho +(1+\beta^2)\omega}\,dt^2  +\bigg(\rho
+(1+\beta^2)\omega\bigg)(\rho+\rho_0)\bigg(d\varphi - \frac{dt} {\rho
+(1+\beta^2)\omega}\bigg)^2 \nn\\ &&+ \frac1{\beta^2\mu_E^2}
\frac{d\rho^2}{\rho^2-\rho_0^2}\,.  \ea This has only one horizon at
$\rho = +\rho_0$, where Kruskal extension can be carried out as
usual. As discussed in \cite{tmgbh} (for $\beta^2=1/4$),  geodesics
actually terminate at the causal singularity $\rho = -\rho_0$, which
is thus a true spacelike singularity of the metric (\ref{sch}). The
resulting Penrose diagram is similar to that of the Schwarzschild
black hole.

Similarly, in the case $\beta^2 = 1$ ($\rho_0=0$), with $r^2$ given
by (\ref{r21}), we find \be \rho_{\pm} = -\omega \pm \sqrt{\omega^2
- 2u}, \ee so that CTCs are absent if  $\omega^2<2u$. The horizon
being double, the corresponding Penrose diagram is similar to that
of the extreme Kerr black hole.  If $u > 0$, CTC do exist but are
hidden behind the double horizon if  $\omega > \sqrt{2u}$. The
Penrose diagram in this case is identical to that of the extreme
Reissner-Nordstr\"{o}m black hole. In the exceptional case $u=0$,
the metric reduces to \be\lb{sch1} ds^2 = -\frac{\rho}{\rho
+2\omega}\,dt^2 + \rho(\rho +2\omega)\bigg(d\varphi - \frac{dt}
{\rho +2\omega}\bigg)^2 +  \frac{d\rho^2}{\mu_E^2\rho^2} \ee (the
$\beta^2=1$, $\rho_0=0$ limit of (\ref{sch})), showing that the
would-be  horizon $\rho=0$ is actually a null singularity.

Finally, in the case $\beta^2 = 0$
(solution(\ref{bh02})-(\ref{r20})), \be c\rho_{\pm} = -\omega \pm
\sqrt{\omega^2 - (\nu+\omega)^2} \ee ($\nu > 0$), so that if
$\omega>-\nu/2$ CTCs are absent. There is a single  horizon at
$\rho=0$, and the reduced two-dimensional ($t,\rho$) metric patches
$\rho>0$ and $\rho<0$ are both asymptotically conformal to Minkowski
2-space.  Accordingly, the Penrose diagram of this geodesically
complete spacetime is  similar to that of the Rindler metric. On the
other hand, if $\omega\le-\nu/2$, $\rho_- > 0$, leading to naked
CTCs.

As in the case of the rotating black holes of TMG \cite{tmgbh}, the
price to pay for the causal regularity of our black holes is that the
Killing vector $\partial_t$ is not timelike, but spacelike (or null
for $\beta^2=1$), so that no observer can remain static outside the
black hole. In other words, the TMGE black holes are surrounded by  an
ergosphere extending from the outer horizon to infinity. However
locally stationary observers are allowed. Their worldline  must remain
timelike ($ds^2<0$), so that for $\rho\ge\rho_0$ fixed,  the angular
velocity $\Omega\equiv d\varphi/dt$ is constrained by  the inequality:
\be r^2\Omega^2-2\Omega(\rho+\omega(1-\beta^2))+ 1 - \beta^2 < 0 \,.
\ee This is satisfied by  \be\lb{obsgen}
\frac{\rho+\omega(1-\beta^2)-\beta\sqrt{\rho^2-{\rho_0}^2}}{r^2}<\Omega<
\frac{\rho+\omega(1-\beta^2)+\beta\sqrt{\rho^2-{\rho_0}^2}}{r^2} \,,
\ee which simplifies in the ``vacuum'' case $\omega=\rho_0=0$ to \be
1-\beta < \Omega\rho < 1+\beta\,.  \ee In the limiting case
$\beta^2=1$ with $\rho_0=0$, the local  stationarity condition
(\ref{obsgen}) becomes \be 0 < \Omega < \frac{2\rho}{r^2}\,.  \ee
Finally, in the special case $\beta^2=0$,  the local stationarity
condition is replaced by \be
\frac{\rho+\omega+\nu-\sqrt{2\nu\rho}}{r^2}<\Omega<
\frac{\rho+\omega+\nu+\sqrt{2\nu\rho}}{r^2} \ee for the generic black
solution (\ref{bh02}), or \be \frac1{\rho+\rho_1} < \Omega <
\frac1{\rho-\rho_1} \ee for the vacuum solution (\ref{vac02}).

\setcounter{equation}{0}
\section{Mass, angular momentum and entropy}

The mass and angular momentum of black hole solutions of TMGE
linearized  about an appropriate background are the Killing charges,
defined as integrals  over the boundary $\part M$ of a spacelike
hypersurface $M$   \be Q(\xi) = \frac1{\k}\int_{\part
M}\sqrt{|g|}{\cal F}^{0i}(\xi)dS_i\,, \ee of the superpotentials
${\cal F}^{\mu\nu}$ associated with the Killing  vectors $\part_t$ and
$\part_{\varphi}$.  These superpotentials may be  written as the sum
\be  {\cal F}^{\mu\nu}(\xi) = {\cal F}^{\mu\nu}_g(\xi) +  {\cal
F}^{\mu\nu}_e(\xi)\,, \ee where the purely gravitational contribution
($A_{\mu} = 0$) is  given in \cite{adtmg} (the sum of Eqs. (2.14) and
(2.21)), and the  electromagnetic contribution is given in \cite{BBCG}
(the sum of Eqs. (68)  and (71)). Rescaling the electromagnetic fields
of \cite{BBCG} by a factor $2\k$, this electromagnetic contribution is
\ba\lb{Fe} {\cal F}^{\mu\nu}_e(\xi) &=&
\frac{\k}{\sqrt{|\hat{g}|}}\delta\bigg[\sqrt{|g|} F^{\mu\nu} -
\mu_E\epsilon^{\mu\nu\lambda}A_{\lambda}\bigg]\xi^{\rho}\hat{A}_{\rho}
\nn \\ && + \k\bigg[\hat{F}^{\mu\nu}\xi^{\rho} +
\hat{F}^{\nu\rho}\xi^{\mu} +  \hat{F}^{\rho\mu}\xi^{\nu}\bigg]\delta
A_{\rho}\,, \ea where the hatted fields are those of the background
(or ``vacuum''), and  $\delta$ stands for the difference between the
fields evaluated for  the black hole configuration and for the
background configuration.  Let us evaluate the radial component ${\cal
F}^{02}_e(\xi)$ in the  adapted coordinates of (\ref{par}) and the
adapted gauge of (\ref{psi}). We recognize in the first bracket of
(\ref{Fe}) the constant of the motion  which was set to zero by the
gauge fixing (\ref{psi}), and there remains  \be {\cal F}^{02}_e(\xi)
=  -\k \hat{F}^{2a}\epsilon_{ab}\xi^b\delta \psi_1 =
-\k\zeta^2(\xi^T\hat\psi)\delta\ol\psi^0\,, \ee which may be
rearranged as \be {\cal F}^{02}_e(\xi) =
\frac{\zeta^2}2\bigg[\xi^T{\bm\Sigma}\cdot \delta{\bm S}_E -
\k(\delta\ol{\psi}\hat\psi)\xi^T\bigg]^0\,.  \ee Combining this with
the gravitational contribution given in  \cite{adtmg}\footnote{The
matrices $\bm\tau$ of \cite{adtmg} are equal to - ${\bm\Sigma}^T$.},
we obtain the net Killing charge for TMGE, \be
Q(\xi)=\frac{\pi\zeta}{\kappa}\left\{\xi^T\bigg({\bm\Sigma}\cdot\delta
{\J}+ \Delta\bigg)\right\}^0\,,   \ee where $\J$ is the constant super
angular momentum \cite{tmge} \be \J = \bL + {\bm S}_G + \S\,, \quad
\bL = \X\wedge\dot\X\,, \quad {\bm S}_G = \lambda\left[\dot\X\wedge\bL
- 2\X\wedge\dot\bL \right]\,, \ee and $\Delta$ is the scalar \be
\Delta = \hat\X\cdot\delta\dot\X+\lambda[\dot{\hat\X}\cdot\delta\bL
-2\hat\X\cdot\delta\dot\bL]-\k(\delta\ol{\psi}\hat\psi)\,.  \ee

The mass and angular momentum are respectively the Killing charges
for the  vectors $\xi = (-1,0)$ and $\xi = (0,1)$: \be\lb{MJ} M =
-\frac{\pi\zeta}{\k}(\delta J^Y + \Delta)\,, \quad J =
\frac{\pi\zeta} {\k}(\delta J^T-\delta J^X)\,.  \ee We shall check
that the values of these observables are consistent with the first
law of black hole thermodynamics,   \be\lb{first} dM = T_{H}dS +
\Omega_{h}dJ\,, \ee where the other observables, readily computed
from the metric in ADM form  \be\lb{adm} ds^2 = -N^2\,dt^2 +
r^2(d\varphi + N^{\varphi}\,dt)^2 + \frac1{(\zeta rN)^2}
\,d\rho^2\,, \ee  are the Hawking temperature and the horizon
angular velocity \be T_{H} = \frac{1}{4\pi}\zeta
r_h(N^2)^{\prime}(\rho_0)\,, \quad \Omega_{h}=
-N^{\varphi}(\rho_{0}) \ee (with $r_h = r(\rho_0)$ the horizon areal
radius), and  the black hole entropy $S$, which is the sum of the
familiar Einstein  contribution and a gravitational Chern-Simons
contribution \cite{solo,tachi, adtmg} \be S  =
\frac{4\pi^2}{\kappa}\bigg(r_h - \lambda r_h^3(N^{\varphi})^{\prime}
(\rho_0)\bigg) \,.   \ee

Let us discuss separately the various cases:

a) In the generic case $0 < \beta^2 < 1$ (solution (\ref{bh2})), the
natural  vacuum is the horizonless metric \be ds^2 = -\beta^2\,dt^2 +
\frac1{\beta^2\mu_E^2}\frac{d\rho^2}{\rho^2} + \rho^2\bigg[d\varphi
-\frac{dt}{\rho}\bigg]^2 \lb{vac}  \ee (the $\omega=0$ member of the
extreme black hole $\rho_0 = 0$ family). The corresponding
observables\footnote{In all cases the two terms $\delta J^Y$  and
$\Delta$ of the r.h.s. of the first equation (\ref{MJ}) contribute
equally  to the mass $M$, which is therefore twice the super angular
momentum value   $-(\pi\zeta/\kappa)\delta J^Y$ as computed in
\cite{tmgbh}.}  \ba\lb{obsg} M &=&
\frac{2\pi\mu_E}{\k}(1-\lambda)\beta^2(1-\beta^2) \omega\,, \nn\\ J
&=& \frac{\pi\mu_E}{\k}\beta^2\bigg[(1-\lambda)(1-\beta^2)\omega^2 -
\frac{1-\lambda(1-2\beta^2)}{1-\beta^2}\rho_0^2 \bigg]\,, \nn\\ S &=&
\frac{4\pi^2}{\k\sqrt{1-\beta^2}} \bigg[(1-\lambda)(1-\beta^2)\omega +
(1-\lambda(1-2\beta^2))\rho_0 \bigg]\,, \nn\\  \Omega_h &=&
\frac{1-\beta^2}{\rho_0+ \omega(1-\beta^2)}\,,\quad  T_H =
\frac{\mu_E}{2\pi}\frac{\beta^2\sqrt{1-\beta^2}\rho_0}{\rho_0+
\omega(1-\beta^2)}  \ea satisfy the first law (\ref{first}) for
independent variations of the black  hole parameters $\omega$ and
$\rho_0$. Let us discuss under what conditions  the mass $M$ and the
entropy $S$ are positive:

($\alpha$) $1/3 \le \lambda<1$ ($\k > 0$). The mass is positive for
$\omega >  0$. The entropy is then positive definite.

($\beta$) $\lambda > 1$ ($\k>0$). The mass is positive for $\omega <
0$, while  the entropy is positive if \be\lb{ubound} \omega <
\frac{1-\lambda(1-2\beta^2)}{(\lambda-1)(1-\beta^2)}\rho_0\,.  \ee
This upper bound for $\omega$ is consistent with the lower bound
ensuring  causal regularity, $\omega > - \rho_0/\sqrt{1-\beta^2}$.

($\gamma$) $-1/3 \le \lambda \le 1/3$ ($\k < 0$). The conditions for
the  positivity of the mass and entropy are the same as in case
($\beta$), however  the upper bound (\ref{ubound}) is not consistent
with causal regularity.

($\delta$) $\lambda < -1/3$ ($\k < 0$). The positivity conditions are
again the same as in case ($\beta$), the upper bound (\ref{ubound})
being conditionally  consistent with causal regularity if $\lambda >
-1$, and always consistent if  $\lambda \le -1$.

b) In the case $\beta^2 = 1$ ($\rho_0=0$), the black hole parameters
are  $\omega$ and $2u$ ($2u > 0$). The vacuum metric (\ref{vac})
corresponds to the parameter values $\omega=2u=0$. As can be expected
from  the limit of the generic case (\ref{obsg}), the values of the
observables  considerably simplify with the mass, horizon angular
velocity and Hawking temperature vanishing, and the angular momentum
and entropy independent of  $\omega$, \ba M &=& 0\,, \quad J =
-\frac{\pi\mu_E}{\k}(1+\lambda)2u\,, \quad S =
\frac{4\pi^2}{\k}(1+\lambda)\sqrt{2u}\,, \nn\\ \Omega_h &=& 0\,,
\qquad T_H = 0\,.  \ea The first law is trivially satisfied. The
entropy is positive either for $\lambda \ge 1/3$ ($\k > 0$) or for
$\lambda < -1$ ($\k < 0$).

c) In the case $\beta^2 = 0$, the black hole parameters are  $\omega$
and $\nu$ ($\nu > 0$) with the vacuum metric (\ref{vac02})
corresponding to the limit $\nu \to 0$. The observables are \ba  M &=&
-\frac{2\pi\mu_E}{\k}(1-\lambda)\nu \,, \quad J =
-\frac{2\pi\mu_E}{\k}\nu\bigg[(1-\lambda)\omega + \nu \bigg]\,, \nn\\
S &=& \frac{4\pi^2}{\k}\bigg[(1-\lambda)\omega + (1+\lambda)\nu
\bigg]\,, \nn\\ \Omega_h &=& \frac{1}{\nu + \omega}\,,\quad  T_H =
\frac{\mu_E}{2\pi}\frac{\nu}{\nu+\omega}\,.  \ea The first law is
satisfied for independent variations of the black hole
parameters. The discussion of the positivity of mass and entropy
parallels  that made in the generic case, with some modifications: in
the case  ($\alpha$) ($1/3 \le \lambda <1$) the mass is now negative
definite, and  in the other cases the upper bound (\ref{ubound})
should be replaced by  \be \omega < \frac{1+\lambda}{\lambda-1}\nu\,.
\ee

d) Finally in the exceptional case $\lambda = 1$ ($\k > 0$), the
metric  generically depends on the three free parameters $\beta^2$,
$\rho_0$ and $\omega$.  However the mass vanishes, so that according
to the first law there should be only one free parameter to vary.
Indeed, for all values of $\beta^2$, we can take  this parameter to
be the entropy $S$, which is positive definite. The other
observables are related to this by \be J =
-\frac{\k\mu_E}{32\pi^3}S^2\,, \quad T_H = \frac{\k\mu_E}{16\pi^3}
\Omega_hS\,,  \ee and the first law is clearly satisfied.

To conclude this section we note that, in all the preceding cases, the
TMGE  black hole observables satisfy, besides the differential first
law  (\ref{first}), the integral Smarr-like relation \be M = T_H S +
2\Omega_h J\,.  \ee This is the natural generalisation of the
Smarr-like relation given in  \cite{black}, Eq. (2.30), where in the
case of TMGE $M$ should be replaced by  $M/2$ in accordance with the
remark made in footnote 2.

\section{Symmetries}
\setcounter{equation}{0}
\subsection{Killing vectors}
The ACL black holes were constructed in \cite{tmgbh} by analytic
extension  of the Vuorio solution, from which they inherited the
local isometry algebra  \linebreak Lie[$SL(2,R)\times U(1)$]
\cite{vuo,ortiz}. This property  generalizes to the case of TMGE
black holes with $\beta^2 > 0$. The metric (\ref{bh2}) admits four
local Killing vectors\footnote{As usual the periodicity condition on
$\varphi$ allows only the two global Killing vectors $\part_t$ and
$\part_{\varphi}$.} $L_t = \part_t$, and $L_0$, $L_{\pm1}$
generating the  $sl(2,R)$ algebra \be [L_m,L_n] = (m-n)L_{m+n}
\qquad  (m,n = -1,0,1)\,, \ee with $[L_t,L_n] = 0$. The $L_n$ are,
in the case of generic $\beta^2 > 0$  black holes, \ba L_0 &=&
\alpha^{-1}(\part_{\varphi}+\omega\part_t)\,, \nn\\ L_{\pm1} &=&
e^{\mp\alpha\varphi}\bigg[\frac1{\alpha\sqrt{\rho^2-\rho_0^2}}
\bigg(\rho(\part_{\varphi}+\omega\part_t) +
\frac{\alpha^2}{\beta^4\mu_E^2}
\part_t\bigg) \pm \sqrt{\rho^2-\rho_0^2}\part_{\rho}\bigg)\bigg]\,,
\ea with \ba\lb{alpha} &\alpha = {\ds \frac{\rho_0}{\gamma}} \;
\left({\ds \gamma = \frac{\sqrt{1-\beta^2}} {\beta^2\mu_E}}\right)&
\quad {\mbox{for}} \quad \beta^2<1,\; (\rho_0\neq0)\,, \nn\\ &\alpha
= \mu_E\sqrt{2u}& \quad {\mbox{for}} \quad \beta^2 = 1,
\;(u\neq0)\,, \ea or, in the case of extreme $\beta^2 > 0$ black
holes ($\beta^2<1$ with $\rho_0=0$ or $\beta^2 = 1$ with $u=0$), \ba
L_1 &=&
\part_{\varphi}+\omega\part_t \,,\nn\\ L_0 &=&
\varphi(\part_{\varphi}+\omega\part_t) - \rho\part_{\rho}\,,\\
L_{-1}  &=& \bigg(\varphi^2 + \frac{\gamma^2}{\rho^2}\bigg)
(\part_{\varphi}+\omega\part_t)  - 2\varphi\rho\part_{\rho} +
\frac{2}{\beta^4\mu_E^2}\rho^{-1}\part_t\,.\nn \ea

In the exceptional case $\beta^2 = 0$, the algebra of local isometries
is  instead the solvable Lie algebra  \ba \left[L_{\pm1}, L_0\right]
&=& \pm L_{\pm1}\,, \nn\\ \left[L_1,L_{-1}\right] &=& -2L_t\,, \\
\left[L_t,L_n\right] &=& 0 \ea ($n = -1,0,1$), for the four local
Killing vectors $L_t = \part_t$ and, in the case of generic black
holes,  \ba L_0 &=& \alpha^{-1}(\part_{\varphi}+\omega\part_t)\,, \nn\\
L_{\pm1} &=& e^{\mp\alpha\varphi}\bigg[\frac1{\sqrt{2\alpha\rho}}
\bigg(\part_{\varphi}+\omega\part_t  - (\rho-\nu)\part_t\bigg) \pm
\sqrt{2\alpha\rho}\part_{\rho}\bigg]\,, \ea with $\alpha = \mu_E\nu
\neq 0$, or, for the vacuum $\nu = 0$, \ba L_0 &=&
\gamma^{-2}\rho(\part_{\varphi}+\omega\part_t) + \gamma^2\varphi
\part_{\rho} + \frac12\bigg(\gamma^{-2}\rho^2 + \gamma^2\varphi^2\bigg)\part_t
\,, \nn\\  L_{\pm1} &=& \gamma^{-1}(\part_{\varphi}+\omega\part_t) \pm
\gamma (\part_{\rho} + \varphi\part_t)\,, \ea with $\gamma^2 =
\mu_E\rho_1^2$.

\subsection{Generating black holes from vacuum}

The fact that the curvature invariants (\ref{curv}) depend only on
$\beta^2$, and that for a given value of $\beta^2$ the local Killing
vectors for our  black holes yield different realizations of the same
Lie algebra suggests  that these solutions may be transformed into
each other by local coordinate transformations. Again we must treat
separately the case of the generic  solution (\ref{bh1}) and the two
special cases $\beta^2 = 0$ and  $\beta^2 = 1$.

a) The metric (\ref{bh1}) may be reduced {\em \`a la} Kaluza-Klein
to an $AdS_2$ ($\varphi,\rho$) metric. This is similar to the case
of the four-dimensional metric $RBR^{-}$ considered in
\cite{conform}, and may be generated from the vacuum metric in the
same manner. Writing (\ref{bh1}) as \be ds^2 =
-\frac{4}{\beta^2\mu_E^2}\frac{(\rho^2-\rho_0^2)}{\rho_0^2}d{u}_+d{u}_-
+ (1-\beta^2)\bigg(d{t} -\omega
d\varphi-\frac{\rho}{1-\beta^2}d\varphi\bigg)^2\,, \ee with \be
{u}_{\pm} = \frac12\bigg(\frac{\rho_0}{\gamma}\varphi \pm
\xi\bigg)\,, \ee where $\gamma$ is defined in (\ref{alpha}), and the
coordinate $\xi$ is related to $\rho$ by \be \rho =
\rho_0\coth\xi\,, \ee we see that it can be obtained from the vacuum
($\omega = \rho_0 = 0$) metric (hatted coordinates) by the combined
transformation \ba \hat{u}_{\pm} &\equiv&
\frac12\bigg(\frac{\rho_0}{\gamma}\hat\varphi \pm \frac{\rho_0}
{\hat\rho}\bigg) = \tanh u_{\pm}\,, \nn\\
\hat{t} &=& t - \omega\varphi + \frac{\gamma}{1-\beta^2}\ln\bigg
(\frac{\cosh u_-}{\cosh u_+}\bigg)\,.
\ea

b) In the case $\beta^2 = 1$, the transformation between the black
hole metric \be ds^2 = \frac1{\beta^2\mu_E^2}\frac{d\rho^2}{\rho^2}
+ (\rho^2+2u)d\varphi^2 -2\rho d\varphi(dt-\omega d\varphi) \ee and
the vacuum metric $\omega = 2u = 0$ is less obvious. However it may
be easily be obtained from the preceding case by putting $\gamma =
\rho_0/\alpha$ and taking the limit $\rho_0 \to 0$ with $\alpha =
\mu_E\sqrt{2u}$ fixed, leading to \ba
\hat\rho &=& \rho\cosh^2\bigg(\frac{\alpha}2\varphi\bigg)\,, \nn\\
\hat\varphi &=& \frac2{\alpha}\tanh\bigg(\frac{\alpha}2\varphi\bigg)\,, \\
\hat{t} &=& t - \omega\varphi
- \frac{\alpha}{\mu_E^2\rho}\tanh\bigg(\frac{\alpha}2\varphi\bigg)\,. \nn
\ea

c) The case $\beta^2 = 0$ (metric (\ref{bh01})) is similar to the
generic case, except that $AdS_2$ is replaced by the flat Rindler
spacetime. Putting $\rho= (\mu_E^2\nu/2)x^2$, (\ref{bh01}) may be
written \be ds^2 = dx^2 - \mu_E^2\nu^2x^2d\varphi^2 + \bigg(dt -
(\omega+\nu)d\varphi - \frac{\mu_E^2\nu}2x^2d\varphi\bigg)^2\,. \ee
This can be obtained from the vacuum metric (\ref{vac01}) (hatted
coordinates) by the combined transformation \ba
\hat\rho &=& \mu_E\rho_1x\cosh(\mu_E\nu\varphi)\,, \nn\\
\hat\varphi &=& \frac{x}{\rho_1}\sinh(\mu_E\nu\varphi)\,, \\
\hat{t} &=& t - (\omega+\nu)\varphi
+ \frac{\mu_Ex^2}4\sinh(2\mu_E\nu\varphi)\,. \nn
\ea

\section{Discussion}

We have constructed intrinsically rotating black hole solutions to
three-dimensional Einstein-Maxwell theory with both gravitational
and electromagnetic Chern-Simons terms. These are geodesically
complete and causally regular within a certain parameter range. We
have computed the mass, angular momentum and entropy of these black
holes, and checked that they satisfy the first law of black hole
thermodynamics. We have also shown that these Chern-Simons black
holes admit a four-parameter local isometry algebra, which
generically is $sl(2,R)\times R$, and that they may be generated
from the corresponding vacua by local coordinate transformations.

As the write-up of this paper was being finalized, we learned about
the paper \cite{ALPSS}, which partly overlaps the present work. This
is concerned with pure TMG, so that our parameter $\beta^2$ is
related to their $\nu^2=-\mu_G^2/9\Lambda$ by $\beta^2 =
(\nu^2+3)/4\nu^2$. Their `spacelike stretched black holes' with
$\nu^2>1$ correspond to our generic $\beta^2<1$ black holes, while
their null warped black hole (6.14) with $\nu^2=1$ corresponds to
our $\beta^2=1$ black hole. They also exhibit local coordinate
transformations generating their $\nu^2>1$ black holes from their
spacelike warped $AdS_3$ (3.3) (which can be obtained from our
solution (3.16) by setting e.g. $c = 1-\beta^2,\, \rho_0^2=-1,\,
\omega=0$, and making the coordinate transformation $\rho =
\sinh\sigma,\, t = \gamma u,\, \varphi = -\gamma\tau$, with $\gamma
= 2\nu\ell/(\nu^2+3)$), and their $\nu^2=1$ black hole from their
null warped $AdS_3$.

This work should be extended in several directions. The close
analogy between our generic black hole metric (\ref{bh1}) and the
rotating Bertotti-Robinson metric $RBR_-$ of \cite{conform} strongly
suggests that the geodesic and test scalar field equations can
similarly be separated and solved in the present case. Likewise, an
investigation of the asymptotic symmetries of our black holes should
extend the local $sl(2,R)\times R$ isometry algebra to a Virasoro
algebra. In another vein, our ansatz (\ref{an}) can be extended to
yield black hole solutions to three-dimensional
Einstein-Maxwell-dilaton theory with an electromagnetic Chern-Simons
term. These shall be reported and discussed elsewhere \cite{tmedbh}.

\end{document}